\documentclass[12pt]{article}
\usepackage{amssymb}
\usepackage{amsfonts}
\usepackage{amsmath,amsthm}
\usepackage{graphics,epsfig}
\usepackage{subfigure}
\usepackage{graphicx,epsfig, color }
\oddsidemargin 10mm \numberwithin{equation}{section}
\def\be{\begin{equation}}
\def\ee{\end{equation}}
\begin{document}
\begin{center}
{{\bf{
 On the stability of electrostatics stars with modified non-gauge invariant Einstein-Maxwell gravity}}
 \vskip 1 cm
 {
H. Ghaffarnejad\footnote{E-mail address:
hghafarnejad$@$semnan.ac.ir}, ~~T. Ghorbani \footnote{E-mail
address: tohidghorbani$@$semnan.ac.ir} and F. Eidizadeh
\footnote{E-mail address: firoozeh.eidizadeh$@$semnan.ac.ir}
}} \\
\vskip 0.1 cm {\textit{Faculty of Physics, Semnan University, P.C.
35131-19111, Semnan, Iran} } \vskip 0.1 cm
\end{center}
\begin{abstract}
We use a modified Einstein-Maxwell gravity to study stability of
an electrostatic spherical star. Correction terms in this model
are scalers which are made from contraction of Ricci tensor and
electromagnetic vector potential. Our motivation to use this kind
of exotic EM gravity is inevitable influence of cosmic magnetic
field in inflation of the universe which is observed now but its
intensity suppresses in the usual gauge invariant EM gravity. In
this work we use dynamical systems approach to obtain stability
conditions of such a star and investigation of affects of
interaction parts of the model on the stability.
\end{abstract}
\section{Introduction}
To describe the stability of a stellar compact object, in usual
way, it is necessary to consider the Tolman-Oppenheimer-Volkoff
equations \cite{1} and the equation of state of the star.
Stability criteria of relativistic spherically symmetric compact
objects with isotropic pressure in the framework of general
relativity include boundary conditions, non-singularity, electric
charge, surface redshift, energy conditions, the speed of sound in
causal conditions and relativistic adiabatic index. In a stable
model, the energy and pressure densities are finite at the center
of compact object and decrease uniformly toward the boundary. The
metric potentials are regular and the electric field intensity is
zero at the center and increases towards the surface. In addition,
the gravitational redshift follows $Z_s <2$  and four energy
conditions are satisfied, the speed of sound is less than the
speed of light and decreases uniformly toward the surface. In
addition, the adiabatic index is strongly higher than
$\frac{4}{3}$ \cite{2}. Relativistic compact objects with gravity
and strong internal density have two different pressures, radial
and tangential \cite{3}. The stability of a stellar model can be
increased by an anisotropic repulsive force that $\Delta =
p_t-p_r> 0$. This property leads to more compact stable
configurations compared to the states of isotropic \cite{4}.
Hydrostatic equilibrium of solutions of anisotropic relativistic
stars in scale-dependent gravity, where Newton's constant is
allowed to vary with radial coordinates across the star, shows
that a decrease in Newton's constant across objects leads to
slightly more massive and compact stars  \cite{5}. A stability
analysis for Einstein-Klein-Gordon model with static real scalar
field interaction express that the initial value of the field at
the origin is a function of the energy density of the matter at
the origin and in the far regions the field behaves Yukawa-like
potential. Such a model for compact stellar object is stable if
the gradient of the total mass versus energy density is positive
and the weak energy condition is satisfied (positive total
density) \cite{6}. The stability of the star can be investigated
in the presence of both electric and magnetic fields. Solving the
Einstein-Maxwell field equations for compact objects with the
charged anisotropic fluid model gives more stable solutions than
for neutral stars. The presence of charges creates a repulsive
force against the gravitational force, and this factor causes
denser stable stars, higher maximum mass and larger redshift
\cite{7}. Charged quarks can create more stable quark stars than
neutron nuclei. Also, for a white dwarf  with a charged perfect
fluid, there is a direct correlation between the increase in
electric charge and its size. Near the surface of the star, the
radial pressure is close to zero and the electric charge density
is non-zero, leading to a stable star with more mass \cite{8}. The
mass-radius relation of some kinds of neutron stars, which can
contain a core of quark matter, has a large frequency range of
radial fluctuations near the transition point in their core versus
mass. These induce nonlinear general relativistic effects which
cause to be the stars unstable dynamically. The core of the
neutron stars becomes several times larger, making the neutron
stars highly unstable \cite{9}. While for the charged
boson-fermion stars with a charged fluid related to fermion and a
complex scalar field related to boson, the charge increase can
reduce the stellar radius and create a denser and more massive
star. In the whole parameter space, the critical curve can show
stable and unstable regions \cite{10}. If the number of baryons in
compact pulsar-like stars exceeds the critical value $10^9$, the
strangeon star model is proposed. In fact the strangeon star
atmosphere  model describes the radiation from interstellar medium
accreted plasma atmosphere on a strangeon star surface and its
spectrum. This object could simply be regarded as the upper layer
of a normal neutron star because the radiation from strangeon
matter can be neglected \cite{11}. The atmosphere is in radiative,
thermal equilibrium and two-temperature. The strangeon star
spectrum is based on bremsstrahlung from an extremely thin
hydrogen plasma. More details of this model are described in
\cite{12}. Since the extra strange flavor provides more degrees of
freedom to lower the Fermi energy in the free quark approximation,
macroscopic bulk strong matter with 3-flavor symmetry (up, down,
and strange quarks) is more stable than up quark matter. The
difference in the strangeness level between a strange star and a
typical neutron star can have a profound effect on the
magnetospheres activity associated with the coherent radio
emission of the compact stars. After to describe several kind of
stellar compact object in summary, we say now about this work and
its content as follows:\\
In section 2 we describe a particular generalized Einstein Maxwell
gravity model which we consider here. In section 3 we obtain Field
equations for a general spherically symmetric static metric. These
are nonlinear second order differential equations and so we use
dynamical systems approach to solve them. We assume that the
electromagnetic source behave same as anisotropic perfect fluid
and generate corresponding density function and radial pressure
and transverse pressure versus the fields. In this section we
generate the Tolman-Oppenheimer-Volkoff equation from conservation
equation of energy tensor field. To solve field equations in the
dynamical systems approach and determine stability of the obtained
solutions one should calculate Jacobi matrix of the set of
differential equations of the system and then determine sign of
its eigenvalues. These are done in the sections 4 and 5 and 6
respectively. The last section dedicated to concluding remarks and
outlook of the work.
\section{The gravity model }
Let us start with the following exotic non-minimally coupled
Einstein Maxwell gravity  \cite{MS}
\begin{equation}\label{action}I=-\int
dx^4\sqrt{g}\bigg[\frac{1}{4}F_{\mu\nu}F^{\mu\nu}+\frac{\alpha}{2}A^2R+\frac{\beta}{2}R_{\mu\nu}A^\mu
A^\nu\bigg],
\end{equation}where $g$ is absolute value of determinant of the metric field and anti symmetric electromagnetic tensor
field $F_{\mu\nu}$ is defined versus the partial derivatives of
the four vector electromagnetic potential $A_{\mu}$ as follows.
\begin{equation} \label{fmunu}
F_{\mu\nu}=\nabla_\mu A_\nu-\nabla_\nu A_\mu=\partial_\mu
A_\nu-\partial_\nu A_\mu
\end{equation} with $A^2=g_{\mu\nu}A^\mu A^\nu$ and $R_{\mu\nu}$ is Ricci
tensor. It is easy to check that this model has not gauge
invariance symmetry same as \cite{GHN} in which the action
functional remain unchanged by transforming $A_{\mu}\to
A_{\mu}+\partial_\mu \xi$ because $F_{\mu\nu}\to F_{\mu\nu}$. In
this transformation $\xi$ is called gauge field.  As we said in
the abstract section this model and other exotic forms of EM
gravity models were presented in the ref. \cite{MS}. Physical
motivation for presentation of these kind of models are influence
of the cosmic magnetic field which is observed now throughout the
universe while it can not be interpreted by ordinary well known
gauge invariant EM gravity. This is because in the cosmic
inflation of the universe the vacuum energy density components
such as quintessence and etc. are dominant terms and density of
cosmic magnetic field is suppress suddenly at duration of
inflation. To keep as non vanishing term in the energy density of
an accelerating expanding universe we must be use other exotic
models such as the above mentioned theory. On the other hand, it
has not been found still in the nature that gauge invariance
symmetry must be maintained in electromagnetic interactions,
therefore we are theoretically free to use models in which gauge
symmetry is broken. Of course, provided that we can reach logical
predictions that are compatible with empirical nature. As an
application of the model (\ref{action}) canonical quantum gravity
approach of this model is studied recently by one of us for a
spherically symmetric electric star and obtained quantum stability
conditions of this kind of stars in ref. \cite{HG}. However we
like to investigate in this work effects of the electromagnetic
fields on stability of an electrostatic stellar object in the
classical approach. To do so we need to solve Einstein metric
equations to obtain internal metric of an electrostatic spherical
perfect fluid by regarding conservation condition of
stress energy tensor. \\
By varying the above action functional
with respect to the electromagnetic vector field $A^\mu$ one can
obtain modified Maxwell equation as
\begin{align}\label{Max}\nabla^{\mu}F_{\mu\nu}=
\frac{\partial^\mu \big(\sqrt{g}F_{\mu \nu}\big)}{\sqrt{g}}=\alpha
R A_\nu + \beta R_{\mu \nu}A^\mu
\end{align}
where right side shows  electric four current which comes from
interaction of gravity and the electromagnetic fields. Also one
can vary the above action functional with respect to the metric
field $g^{\mu\nu}$ to obtain modified Einstein metric field
equation such that
\begin{align}\label{G}
G_{\mu\nu}&=\frac {2} {\alpha A^2}\bigg(\xi T_{\mu\nu}-\frac{
T_{\mu\nu}^{EM}}{2}- \frac{\alpha}{2} A_{\mu}A_{\nu} R+
\frac{\beta}{4} g_{\mu\nu}
R_{\kappa\lambda}A^{\kappa}A^{\lambda}\notag\\&-\frac{1}{4}(
\nabla_\lambda \nabla_\nu \theta_\mu ^\lambda+\nabla_\lambda
\nabla_\mu \theta_\nu ^\lambda)+ \frac{1}{4} g_{\mu\nu}
\partial_\kappa \big(\nabla_\lambda \theta^{\kappa\lambda}\big)+\frac{1}{4}
\Box \theta_{\mu\nu}\bigg)
\end{align}
where $T_{\mu\nu}$ is  additional arbitrary other sources which we
will dropped in what follows and the vacuum sector of the EM field
stress energy tensor is
\begin{equation}T^{EM}_{\mu\nu}=\frac{1}{2}\bigg(F_{\mu\kappa}F_{\nu}^\kappa+F_{\nu\lambda}F_{\mu}^{\lambda}\bigg)-\frac{1}{4}g_{\mu\nu}F^2,~~
~F^2=F_{\zeta\sigma}F^{\zeta\sigma}\end{equation} and also we
defined
\begin{equation}\theta_{\mu\nu}=\alpha A^2g_{\mu\nu}+\beta A_{\mu}A_{\nu}.\end{equation} Covariant conservation of total
matter stress energy tensor (right side in eq. (\ref{G})) or
equivalently Bianchi identity ($\nabla^{\mu}G_{\mu\nu}=0$) reduce
to Tolman-Oppenheimer-Volkoff equation which we present in the
subsequent section. In fact, this equation describes variation of
radial pressure $p_r(r)$ of the stellar fluid versus the matter
density $\rho(r)$ and transverse pressure $p_t(r)$ and mass
function $m(r)$ of the stellar fluid, if  total matter stress
tensor in right side of the equation (\ref{G}) behaves same as
anisotropic spherically symmetric perfect fluid
 such that
\begin{equation} (T^{\mu}_{\nu})_{total}=diag[\rho(r), p_r(r),p_t(r),p_t(r)]\end{equation} in which $r$ is a radial coordinate
in a local spherically symmetric coordinates system and the metric
field equation (\ref{G}) reads to a simplest form as
\begin{equation}\label{G2}G^\mu_\nu=(T^{\mu}_{\nu})_{total}\equiv\left(%
\begin{array}{cccc}
  \rho(r) & 0 & 0 & 0 \\
  0 & p_r(r) & 0 & 0 \\
  0 &0 & p_t(r) & 0\\
  0 & 0 & 0 & p_t(r) \\
\end{array}%
\right)\end{equation} where we suppressed  the factor $8\pi G.$
\section{Tolman-Oppenheimer-Volkoff equation}
For spherically symmetric time-independent static metric
\begin{equation}\label{line}ds^2=U(r)dt^2-V(r)dr^2-r^2(d\theta^2+\sin^2\theta
d\varphi^2)\end{equation} it is easy to check that
\begin{equation}\label{at}A_\mu(r)=\phi(r)\delta_{\mu t}\end{equation} is only
non-vanishing component of the electromagnetic vector potential in
which $\delta$ is Kronecker delta function. By substituting
(\ref{at}) into the Maxwell equation (\ref{Max}) together with the
line element (\ref{line}) we obtain
\begin{align}\label{33}\psi^\prime &+(\beta+2\alpha)f^\prime \notag\\&+(\frac{\psi}{2}+\frac{2\alpha}{r}+\frac{(\beta-2\alpha)f}{2})\frac{V^\prime}{V}\notag \\&=
\psi
f-\frac{2\psi}{r}-\psi^2+(\beta-6\alpha)f^2-\frac{2(\beta+2\alpha)f}{r}-\frac{2\alpha(V-1)}{r^2}\end{align}
 where $\prime$ is derivative versus the $r$ coordinate   and we defined \begin{equation}
 \label{def}\psi(r)=\frac{\phi^\prime}{\phi},~~~f(r)=-\frac{1}{2}\frac{U^\prime}{U}.\end{equation}
 By
comparing (\ref{G}) and  (\ref{G2}) and by substituting
(\ref{line}), (\ref{at}) and (\ref{def}) one can find
\begin{align}\label{36}\rho(r)&=\frac{(\beta-8\alpha)}{2\alpha}\frac{f^\prime}{V} -(\beta+2\alpha)\frac{\psi^\prime}{V}
+\bigg[\frac{(\beta V+3\alpha)}{2\alpha}\psi
+\frac{(10\alpha-\beta)f}{4\alpha} \notag \\&-\frac{2}{r
}\bigg]\frac{V^\prime}{V^2}+\frac{2(1-V)}{r^2
V}-3(\beta+3\alpha)\frac{\psi
f}{V}-\frac{2\beta}{\alpha}\frac{\psi}{rV} \notag
\\&+\frac{(1-4\beta-8\alpha)}{2\alpha}\frac{\psi^2}{V}
-\frac{(4\alpha+7\beta)}{2\alpha}\frac{f^2}{V}+\frac{(\beta-6\alpha)}{\alpha}\frac{f}{rV}\end{align}
\begin{align}\label{37}p_r(r)&=\frac{\beta}{2\alpha}\frac{f^\prime}{V}+\bigg[\frac{\psi}{2}+\bigg(1-\frac{\beta}{2\alpha}\bigg)\frac{f}{2}\bigg]
\frac{V^\prime}{V^2}+\bigg(1+\frac{5\beta}{2\alpha}\bigg)\frac{f^2}{V}
\notag\\&
+\bigg(1+\frac{2\beta}{\alpha}\bigg)\frac{f\psi}{V}-\frac{2\psi}{rV}+\bigg(\frac{\beta}{\alpha}-2\bigg)\frac{f}{rV}+\frac{\psi^2}{2\alpha
V}\end{align}
\begin{align}\label{38}p_t(r)&=\frac{(\beta-4\alpha)}{2\alpha}\frac{f^\prime}{V}+\bigg[\frac{3}{2}\psi+\frac{(12\alpha-\beta)f}{4\alpha}
\bigg]\frac{V^\prime}{V^2} \notag\\&
-\frac{(1+8\alpha)}{2\alpha}\frac{\psi^2}{V}-\frac{(6\alpha+\beta)}{2\alpha}\frac{f^2}{V}+\frac{\beta
f}{\alpha rV}-\frac{7f\psi}{V}.\end{align} For spherically
symmetric line element same as (\ref{line}) even if to be
non-static with $(t,r)$ dependency, one can show that the Maxwell
tensor field $F_{\mu\nu}(t,r)$ can be rewritten versus the polar
components of the electric $\overrightarrow{E}(t,r)$ and magnetic
$\overrightarrow{B}(t,r)$ fields such that
\begin{equation}\label{Fmunu}F_{\mu\nu}
 =\left(%
\begin{array}{cccc}
  0 &-E_r & -rE_\theta & -r\sin\theta E_\varphi \\
   E_r& 0 & rB_\varphi & -r\sin\theta B_{\theta} \\
   rE_\theta  & -rB_\varphi & 0 & r^2\sin\theta B_r \\
 r\sin\theta E_\varphi & r\sin\theta B_\theta & -r^2\sin\theta B_r& 0 \\
\end{array}%
\right).
\end{equation}
In the static form of the metric field where the electromagnetic
fields should be only $r$ dependent then all components of the
above matrix vanish except radial electric component $E_r(r)$
given by (\ref{at}). In fact covariant conservation condition of
the matter stress tensor or Bianchi`s identity (\ref{G2}) gives a
relation between the  matter density and pressures together with
the mass function (in unites $c=G=1$)
\begin{equation}\label{mr}m(r)=4\pi\int_0^r\rho(\bar{r})\bar{r}^2d\bar{r}
\end{equation}
which is called  Tolman Oppenheimer-Volkoff (TOV) equation such
that
\begin{equation}\label{p}p^\prime_r=-(\rho+p_r)f+\frac{2(p_t-p_r)}{r}\end{equation}
where we use the following ansatz for the metric components of the
line element (\ref{line}).
\begin{equation}\label{V}V(r)=\bigg(1-\frac{2m(r)}{r}\bigg)^{-1}\end{equation}
and
\begin{equation}\label{U}U(r)=\exp\bigg(-2\int_r^\infty f(\bar{r})d\bar{r}\bigg)\end{equation}
in which
\begin{equation}\label{g}f(r)=\frac{m(r)+4\pi r^3p_r(r)}{r[r-2m(r)]}\end{equation} is the locally measured gravitational
acceleration, and is pointing inwards for positive gravitational
acceleration $g(r)$ (see \cite{Cel} and \cite{Hob} for more
details).
 The TOV equation describes variation of radial pressure of an anisotropic stellar compact fluid versus its density and transverse pressure  and
the metric field. To solve the TOV equation we need another
equation which relates pressures to the density function. It is
called equation of state and come from statistical distribution of
fundamental particles which make the stellar fluid. By having a
known equation of state and some suitable boundary conditions for
stellar object one can solve the above TOV equation and then find
the interior metric solutions (\ref{U}) and (\ref{V}).  Regardless
of the anisotropies, it is well known that for many astrophysical
systems the matter satisfies a polytropic form of equation of
state as $p_r=p_t=p=K\rho^{1+\frac{1}{n}}$ where $K$ and $n$ are
constants related to relative heat capacities as
$1+\frac{1}{n}=\frac{C_p}{C_V}$ in which $C_p$ and $C_V$ are heat
capacities at constant pressure $p$ and volume $V$. $n$ is so
called the polytropic index. In particular case $n\to\infty$ this
equation of state reads $p=K\rho$ in which $K$ is so called the
barotropic index and the star is in isothermal state while for
$n=0$ density of the star under consideration is constant and for
$n=1.5$ the star is in convection equilibrium. In the model under
consideration the dynamical field equations have not simple forms
and so we can not obtain exact analytic solutions. Hence we must
be use approximation methods. One of these methods is use of
dynamical systems approach and obtain some analytic exact
solutions around some assumed critical points. To do so we should
first obtain closed form of the equations such as follows.
\\
By solving the equations (\ref{33}), (\ref{36}), (\ref{37}) and
(\ref{38}) versus $\psi^\prime,f^\prime,V^\prime$ and $p_t$ one
can obtain
\begin{align}\label{vp}V^\prime&=\frac{-2V\Sigma_V(f,V,\psi, \rho, p_r;r)}{r\Sigma(f,V,\psi;r)},\notag\\&~~~f^\prime=\frac{\Sigma_f(f,V,\psi, \rho, p_r;r)
}{r\Sigma(f,V,\psi;r)},\notag\\&~~~\psi^\prime=
-\frac{\Sigma_\psi(f,V,\psi, \rho,
p_r;r)}{r^2\Sigma(f,V,\psi;r)},\end{align} and
\begin{equation} p_t=-
\frac{\Sigma_{p_t}(f,V,\psi, \rho, p_r;r)}{\alpha
rV\Sigma(f,V,\psi;r)}\end{equation} where definitions of the
functions $\Sigma$ and $\Sigma_{V,f,\psi,p_t}$ are given in the
appendix I. The equations (\ref{vp}) together with the TOV
equation (\ref{p}) are all dynamical equations which determine
interior metric of a compact stellar object. They are set of
nonlinear first order differential equations and can be solve via
dynamical system approach. To do so we first make them as closed
form which means that each derivative functions in (\ref{vp}) and
(\ref{p}) should defined just with fields $V,f,\psi,p_r.$ This is
done by defining equation of states $p_r(\rho)$ and $p_t(\rho)$
where we use the ansatz $p_r=K\rho^{1+\frac{1}{n}}$ for radial
pressure. Also by looking at the critical points obtained at below
section we infer that it is better to choose $p_t=\gamma \rho$ for
transverse pressure.
\section{Critical points}
 In the dynamical systems approach the critical
points \begin{equation}\{V_c,f_c,\psi_c,p_{rc},p_{tc},
\rho_c\}\equiv constant\end{equation} are obtained by solving the
equations
\begin{equation}\label{crit}V^\prime=0=f^\prime=\psi^\prime=p_r^\prime\end{equation} for each
critical radius $r_c.$ To obtain physical solutions of the above
dynamical equations around given critical points we must be use
some physical initial conditions. To have interior metric of a
compact stellar object we assume $r_c=R$ to be radius of a compact
stellar object with total mass $M=m_c=m(r_c)$ in which the
critical radial pressure must be vanish at the stable state namely
$p_{rc}=0$ while the transverse pressure may not be vanish
$p_{tc}\neq0.$ Hence we choose the following ansatz for initial
state of stellar compact object
\begin{equation}\label{crit2}p_{rc}(R)=0,~~~\{V_c(R),f_c(R),\psi_c(R),p_{tc}(R),
\rho_c(R)\}\neq0.\end{equation} By regarding the above initial
condition on the surface of a compact stellar object the equations
(\ref{V}) and (\ref{g}) give us
\begin{equation}\label{crit3}V_c(R)=1+2Rf_c(R),\end{equation} and \begin{equation}
\frac{2M}{R}=\frac{2Rf_c(R)}{1+2Rf_c(R)}<1\end{equation} which
means that we have a star at the critical radius $R$ which is
larger than the corresponding Schwarzschild radius $2M$. By
substituting these relations into the critical equations
(\ref{crit}) we obtain
\begin{equation}\label{deff}Rf_c(R)=\frac{2p_{tc}(R)}{\rho_c(R)}=2\gamma,~~~\psi_c=\frac{\sigma}{R},~~~\rho_c=\frac{\eta}{R^2}\end{equation}
where \begin{align}\label{def2}\alpha
&=\frac{\sigma(5\gamma^2-\gamma\sigma-\sigma^2-4\gamma-2\sigma-1)}{2(32\gamma^3+13\gamma^2\sigma+22\gamma^2+6\gamma\sigma+6\gamma+\sigma)}\notag\\&
\beta=-\frac{\sigma(4\gamma^3+6\gamma^2\sigma-\gamma\sigma^2-8\gamma^2+2\gamma\sigma+\sigma^2+4\gamma+2\sigma)}{4\gamma(32\gamma^3+13\gamma^2\sigma
+22\gamma^2+6\gamma\sigma+6\gamma+\sigma)}\notag\\&
\eta=-\frac{2}{\gamma}(28\gamma^4+42\gamma^3\sigma+4\gamma^2\sigma^2-9\gamma\sigma^3-2\sigma^4\notag\\&
-18\gamma^3-27\gamma^2\sigma-20\gamma\sigma^2
-4\sigma^3-12\gamma^2-4\gamma\sigma-\sigma^2+2\gamma+\sigma
)/(20\gamma^3\notag\\&-4\gamma^2\sigma-4\gamma\sigma^2-11\gamma^2-9\gamma\sigma-\sigma^2-8\gamma-2\sigma-1)\end{align}
 and the two parameters $\gamma$ and $\sigma\neq0$ satisfy the
 following equation.
\begin{align}\label{def3}&(15\gamma^2+2\gamma+1)\sigma^6-(27\gamma^3-135\gamma^2-42\gamma-10)\sigma^5\notag \\&
-(569\gamma^4-389\gamma^3-519\gamma^2-148\gamma-17)\sigma^4\notag
\\&
-(846\gamma^5+1085\gamma^4-1491\gamma^3-177\gamma-895\gamma^2-6)\sigma^3\notag
\\&-
(1614\gamma^6+7314\gamma^5+750\gamma^4-624\gamma^2-1680\gamma^3-18\gamma
+4)\sigma^2\notag\\&-(4936\gamma^7+12416\gamma^6+6552\gamma^5-400\gamma^4-728\gamma^3-8\gamma^2+32\gamma)\sigma\notag\\&
+768\gamma^8-240\gamma^7-1408\gamma^6+432\gamma^5+432\gamma^4+64\gamma^3-48\gamma^2=0.\end{align}
 By looking at the
equation (\ref{deff}) we see that at critical point
$p_{tc}=\gamma\rho_c$ and so we are allowed to use
\begin{equation}\label{pt}p_t=\gamma \rho
\end{equation} for transverse part of equation of state  which we pointed at the previous section. For radial part of equation of state
$p_r(\rho)=K\rho^{1+\frac{1}{n}}$ is useful to write versus a
dimensionless parameter function $y(x)$ such that
\begin{equation}p_r=p_{0r}y^{n+1}(x),~~~\rho=\rho_0y^n(x),~~~x=\frac{r}{R},~~~K=\frac{p_{0r}}{\rho_0^{1+\frac{1}{n}}}\end{equation} where
$R$ is radius of the compact star and $\rho_0$ and $p_{0r}$ are central density and radial pressure respectively.
With this definition one can show that the transverse equation of
state  can be rewritten as follows.
\begin{equation}p_t=\gamma\rho_0y^n(x).\end{equation} By substituting these definitions into the TOV
equation (\ref{p}) we can rewrite it versus $y(x)$ as follows.
\begin{equation}\label{ydot}
\dot{y}=-\frac{u(x)}{(1+n)}\bigg(\frac{1}{\delta}+y\bigg)+\frac{2}{(1+n)}\bigg(\frac{\gamma}{\delta}-y\bigg)\frac{1}{x}\end{equation}
where we defied dimensionless quantities
\begin{equation}\delta=\frac{p_{0r}}{\rho_0}=K\rho_0^\frac{1}{n},~~~u(x)=Rf(xR)\equiv Rf(x)\end{equation} and $\dot{~}$ is derivative with respect to
$x.$ It is useful to write a dimensionless form for the
differential equations (\ref{vp}) by defining
\begin{equation}z=R\psi(x)\end{equation}
which in small scales limits $0<x<1(\equiv r<R)$ reads to the
following forms.
\begin{align}\label{r0}\dot{V} &\approx\frac{h_1}{x}V(1-V)\\\notag &
\dot{u}\approx\frac{h_2}{x}z
V+\frac{h_3}{x}z+\frac{h_4}{x}uV+\frac{h_5}{x}u\\\notag &
\dot{z}\approx-\frac{h_6}{x^2}(1-V)\end{align} where we defined
\begin{align}&h_1=\frac{4\alpha^2+2\alpha\beta-1}{2(2\alpha^2+\alpha\beta-1)},~~~h_2=\frac{\alpha(4\alpha^2+2\alpha\beta-1)}{2\beta(2\alpha^2+\alpha\beta-1)}\\\notag &
h_3=\frac{\alpha(12\alpha^2+6\alpha\beta-7)}{2\beta(2\alpha^2+\alpha\beta-1)},~~~h_4=\frac{(2\alpha-\beta)[2\alpha(2\alpha+\beta)-1]}{4\beta(2\alpha^2+\alpha\beta-1)}\\\notag&
h_5=\frac{(2\alpha-\beta)[6\alpha(2\alpha+\beta)-7]}{4\beta(2\alpha^2+\alpha\beta-1)},~~~h_6=\frac{\alpha}{(2\alpha^2+\alpha\beta-1)}.\end{align}
The equations (\ref{r0}) and the TOV equation (\ref{ydot}) have
closed form which means each derivative function can be described
just with other fields and there is not every extra field in right
side of these equations with no derivative function. This closed
form make a 4D phase space $\{V, u,z,y\}.$ In the subsequent
section we apply to solve these equations via dynamical systems
approach and investigate stability conditions of the obtained
metric solutions. \section{Metric solutions} At first step in the
dynamical systems approach, we should linearized the set of
differential equations (\ref{ydot}) and(\ref{r0}) by calculating
the Jacobi matrix $J_{ij}=\frac{\partial \dot{X}_i}{\partial
X_j}$. In the dynamical systems approach each set of nonlinear
first order differential equations with closed form can be
linearized versus the Jacobi matrix and the fields as
$\dot{X}_i=\Sigma_{j=1}^nJ_{ij}X_j$ where $i,j=1,2,3,\cdots n$ for
$n$ dimensional phase space of the system and $J_{ij}$ should be
calculated at critical points. For set of the equations (\ref{r0})
and (\ref{ydot}) we obtain the following critical points by
solving the equations $\dot{y}=0=\dot{V}=\dot{u}=\dot{z}.$
\begin{equation}V_c=1,~~~y_c=\frac{1}{\delta}\bigg(\frac{2\gamma-u_c}{2+u_c}\bigg),~~~z_c=-\bigg(\frac{h_4+h_5}{h_2+h_3}\bigg)u_c,~~~x_c=1.
\end{equation} These critical points are parametric and can be fixed by physical boundary condition as follows. It is important to note that
on the star surface $x_c=1$ the radial pressure vanishes $y_c=0$
which by substituting these into the above parametric critical
points we obtain finally
\begin{equation}\label{cf}c.p:~~~\{V_c=1,~~~y_c=0,~~~u_c=2\gamma,~~z_c=-2\gamma\bigg(\frac{h_4+h_5}{h_2+h_3}\bigg)\}_{|_{x_c=1}}.\end{equation}
One can use this critical point to calculate $J_{ij}$ such that
\begin{equation}\label{Jac}{J_{ij}}_{\big|_{c.p}}=\left(%
\begin{array}{cccc}
  J_{11} & 0 & J_{13} &0\\
  0& J_{22} & 0 &0\\
  0 & J_{32} & J_{33} &J_{34}\\
0&J_{42}&0&0\end{array}%
\right) \\
\end{equation}
where \begin{align}&J_{11}=\frac{\partial \dot{y}}{\partial
y}{\Big|_{c.p}}=-2\bigg(\frac{1+\gamma}{1+n}\bigg),~~~
J_{13}=\frac{\partial \dot{y}}{\partial
u}{\Big|_{c.p}}=\frac{-1}{\delta(1+n)}\\\notag&
J_{22}=\frac{\partial \dot{V}}{\partial
V}{\Big|_{c.p}}=-h_1,~~~J_{32}=\frac{\partial \dot{u}}{\partial
V}{\Big|_{c.p}}=2\gamma\bigg(\frac{h_3h_4-h_2h_5}{h_2+h_3}\bigg)\\\notag&J_{33}=\frac{\partial
\dot{u}}{\partial u}{\Big|_{c.p}}=h_4+h_5,~~~J_{34}=\frac{\partial
\dot{u}}{\partial z}{\Big|_{c.p}}=h_2+h_3,~~~J_{42}=\frac{\partial
\dot{z}}{\partial V}{\Big|_{c.p}}=h_6.\end{align} Near the
critical point (\ref{cf}) the dynamical equations
 can be written as $\dot{X}_i=\Sigma_{j=1}^nJ_{ij}X_j$ such that
\begin{equation}\frac{d}{dx}\left(%
\begin{array}{c}
  y \\
  V \\
  u \\
  z \\
\end{array}%
\right)=\left(%
\begin{array}{cccc}
  J_{11} & 0 & J_{13} &0\\
  0& J_{22} & 0 &0\\
  0 & J_{32} & J_{33} &J_{34}\\
0&J_{42}&0&0\end{array}%
\right) \\\left(%
\begin{array}{c}
  y \\
  V \\
  u \\
  z \\
\end{array}%
\right)\end{equation} which have solutions as follows.
\begin{align}&V(x)=e^{J_{22}(x-1)}\\\notag&
z(x)=z_c+\frac{J_{42}}{J_{22}}\bigg[e^{J_{22}(x-1)}-1\bigg]\\\notag&
u(x)=\frac{(J_{32}J_{22}+J_{34}J_{42})}{J_{22}(J_{22}-J_{33})}\big[e^{J_{22}(x-1)}-e^{J_{33}(x-1)}\big]\\\notag&+\frac{J_{34}}{J_{33}}\bigg(z_c-\frac{J_{42}}{J_{22}}
\bigg)\big[e^{J_{33}(x-1)}-1\big]+u_ce^{J_{33}(x-1)}\\\notag&
y(x)=A[1-e^{J_{11}(x-1)}]+B[e^{J_{11}(x-1)}-e^{J_{22}(x-1)}]+C[e^{J_{33}(x-1)}-e^{J_{11}(x-1)}]\end{align}
where we defined
\begin{align}&A=\frac{J_{13}J_{34}}{J_{11}J_{33}}\bigg(z_c-\frac{J_{42}}{J_{22}}\bigg)\\\notag&B=\frac{J_{13}(J_{22}J_{32}+J_{34}J_{42})}{J_{22}(J_{11}-J_{22})(J_{22}-J_{33}
)}\\\notag&C=\frac{J_{13}}{(J_{11}-J_{33})}\bigg[\frac{J_{32}J_{22}+J_{34}J_{42}}{J_{22}(J_{22}-J_{33})}-\frac{J_{34}}{J_{33}}\bigg(z_c-\frac{J_{42}}{J_{22}}\bigg)-u_c\bigg].\end{align}
Using the above solutions one can show that
\begin{equation}\ln\bigg(\frac{\phi}{\phi_c}\bigg)=\bigg(z_c-\frac{J_{42}}{J_{22}}\bigg)(x-1)+\frac{J_{42}}{J_{22}^2}\bigg[e^{J_{22}(x-1)}-1\bigg]
\end{equation}
with dimensionless electric field
\begin{equation}\bar{E}_r=\frac{RE_r}{\phi_c}=z(x)\exp\bigg\{\bigg(z_c-\frac{J_{42}}{J_{22}}\bigg)(x-1)+\frac{J_{42}}{J_{22}^2}[e^{J_{22}(x-1)}-1]
\bigg\}\end{equation} in which we defined $\phi_c=\phi(x=1)$ and
\begin{equation}U(r)=e^{-2\int u(x)dx},~~~\rho=\rho_0y^n,~~~p_r=p_{0r}y^{n+1},~~~p_t=\gamma\rho_0y^n\end{equation}
with mass-radius relation
\begin{equation}\frac{2m}{r}=\frac{2xu(x)-x^2 y^{n+1}(x)}{2xu(x)+1},~~~8\pi R^2p_{0r}=1\end{equation}
which is obtained from the equation (\ref{g}).
 To determine stability conditions of the above
obtained solutions we must be solve secular equation of the above
jacobi matrix defined by $\det(J_{ij}-\varepsilon\delta_{ij})=0$
and determine sign of the eigenvalues $\varepsilon.$ If four
eigenvalues take real negative (positive) sign then the obtained
solutions become stable (unstable). If they become complex numbers
with negative (positive) sign for the real part of complex
eigenvalues then nature of the obtained solutions will be spiral
stable (unstable) state. In case where some of eigenvalues are
zero  then the system will be degenerate and stability/instability
of its future are dependent to effects of other external
perturbations forces (see introduction section of ref. \cite{15}
for more discussions about the dynamical systems approach). In the
next section we analyzes the eigenvalues of the system under
consideration  as follows.
\section{Eigenvalues}
It is easy to show that the secular equation
$\det(J_{ij}-\varepsilon\delta_{ij})=0$ for the Jacobi matrix
(\ref{Jac}) reads
\begin{equation}\varepsilon(J_{11}-\varepsilon)(J_{22}-\varepsilon)(J_{33}-\varepsilon)=0\end{equation}
which has solutions
\begin{align}&\varepsilon_1=J_{11}=-2\bigg(\frac{1+\gamma}{1+n}\bigg)\\\notag&
\varepsilon_2=J_{22}=-h_1=\frac{-1}{2}\bigg(\frac{4\alpha^2+2\alpha\beta-1}{2\alpha^2+\alpha\beta-1}\bigg)=\frac{-1}{1-1/(4\alpha^2+2\alpha\beta-1)}\\\notag&
\varepsilon_3=J_{33}=h_4+h_5=2\bigg(\frac{2\alpha}{\beta}-1\bigg)\\\notag&\varepsilon_4=0.\end{align}
The presence of a zero root $\varepsilon_4=0$ indicates that the
system is degenerated at all and by adding some other perturbation
sources may reaches to stable or unstable states and we will
investigate this case as our future work. Regardless to this zero
eigenvalue which can be resolve by considering a time evolution of
the collapsing compact stellar object, we must set
$\varepsilon_{1,2,3}<0$ for which $\varepsilon_3$ gives us
$2\alpha<\beta$ and $\varepsilon_2$ gives us
$4\alpha^2+2\alpha\beta-1>1.$ These inequalities are shown in
figure 1 for permissable values of the parameters $\alpha,\beta$
which make $\varepsilon_{2,3}<0.$ For $\varepsilon_1<0$ we know
that $n>0$ and so we must be choose $\gamma>-1.$ For instance if
we set $\gamma=-\frac{1}{3}$ then the barotropic index for
transverse pressure behaves as quintessence dark energy while for
$\gamma=-\frac{2}{3}$ this behaves as fantom phase of the dark
energy. For super fluid (supper sonic) the sound speed reaches to
a maximum value such that $\gamma=1$ which we consider here such
that $\varepsilon_1=\frac{-4}{1+n}.$ In the latter case we have
degenerate state $\varepsilon_0=0$ for isothermal star with
$n\to\infty$ and for constant density with $n=0$ we have
$\varepsilon_1=-4$ while for star in convection equilibrium with
$n=1.5$ we have $\varepsilon_1=-1.6.$ Furthermore in central
region of a compact stellar objects we can consider the fluid
behaves as isotropic and homogenous namely
\begin{equation}\delta=\frac{p_{0r}}{\rho_0}=\gamma=\frac{p_{0t}}{\rho_0}.\end{equation} In summary, by looking at the figure 1 we
choose anstaz
\begin{equation}\alpha=1,~~~\beta=3,~~~\gamma=\delta=1\end{equation} for
numerical studies in what follows and obtain
\begin{equation}\varepsilon_1=\frac{-4}{1+n},~~~\varepsilon_2=-\frac{9}{8},~~~\varepsilon_3=-\frac{2}{3},~~~n>0\end{equation}
and
\begin{align}& h_1=\frac{9}{8},~~~h_2=\frac{3}{8},~~~h_3=\frac{22}{24},~~~h_4=-\frac{3}{16},~~~h_5=-\frac{23}{48},~~~h_6=\frac{1}{4}\\\notag&
u_c=2,~~~z_c=1,~~~J_{11}=\frac{-1}{1+n},~~~J_{22}=-\frac{9}{8},~~~J_{33}=-\frac{2}{3},~~~J_{13}=\frac{-1}{1+n},\\\notag&
J_{32}=0,~~~J_{34}=\frac{4}{3},~~~J_{42}=\frac{1}{4},~~~
A=-\frac{11}{18},~~~B=\frac{512}{99(9n-23)}\\\notag&C=\frac{18}{11(5-n)}.\end{align}
For these numeric values we plotted dimensionless radial electric
field $\bar{E}_r(x),$ dimensionless matter density
$\bar{\rho}=\frac{\rho(x)}{\rho_0}$, dimensionless pressures
$\bar{p}_r=\frac{p_r}{p_{0r}}$ and $\bar{p}_t=\frac{p_t}{p_{0t}}$
and mass per radius relation $\frac{2m}{r}$ in figures 2 for
different values of the polytropic index $n$ parameter. By looking
at the figure 2-a one can infer that by rasing the radial distance
the electric field intensity increases and take on its maximum
value on the surface of star. While internal metric components
decreases. The figure 2-b shows that decreasing slope of the
density function decreases faster by raising he radial distance of
the star from its center and vanishes on the star surface. There
is similar behavior for the transverse pressure and radial
pressure but with larger scale. For smallest value of the
polytropic index $n$ slope of density diagram by raising the
radial distance of the star is very slow but it is dropped
suddenly near the star radius. The figure 2-d shows variations of
the mass per radius relation of the compact star with positive
slope such that its maximum value does not reach to Schwarzschild
radius means that our obtained stellar object is really an visible
star and not a black hole. In summary, by looking at these
diagrams one can infer that the obtained solutions describe a
electrostatic spherically symmetric anisotropic star with maximal
stability at classical regimes of the field. This results obey
results of the quantum regimes of the field given in the ref.
\cite{HG} which is investigated recently by one of us. In the next
section discuss outputs of the work and future ideas for extension
of the work.
\section{Concluding remarks}
In this work we added a nonminimal directionally interaction
Lagrangian between geometry and the electromagnetic vector
potential for Einstein-Maxwell gravity and investigated  this
additional contribution on internal space time of spherically
symmetric static stellar compact object. After to solve the
Euler-Lagrange equations of the fields via dynamical systems
approach, we determined stabilization conditions of the obtained
solutions near parametric critical points in phase space. We
obtained permissable numeric values of the parameters of the
interaction Lagrangian parts which give stable nature for the
obtained solutions. This results are found by determining sign of
eigenvalues of the Jacobi matrix of the dynamical equations of the
system. One of the four eigenvalues is zero value while other tree
eigenvalues were parametric which by choosing  suitable numeric
values for the parameters they become negative sign. However in
the dynamical system approach the system become full stable if all
eigenvalues become negative real numbers. If one of the is zero
then the system become quasi stable. Hence to make negative values
for zero eigenvalue we should consider other sources which can be
break this degeneracy. This will done in our future work by
considering the magnetic field (see \cite{GHN} for magnetic
monopole application). However by choosing a polytropic form of
the equation of state we show that the stability of the system is
dependent to particular values of the polytropic index of the
system together with the two coupling constant of the gravity
model under consideration.

\section{Appendix I}
\begin{align}\Sigma&=8\alpha^3\beta+4\alpha^2\beta^2-4\alpha\beta\notag
\\&r[(8\alpha^2-8\alpha^4-8\alpha^3\beta+2\alpha^2\beta^2+2\alpha\beta^3)f\notag\\&
+(2\alpha\beta-8\alpha^4-8\alpha^3\beta-2\alpha^2\beta^2+V\beta^2+2\alpha^2\beta+\alpha\beta^2+8\alpha^2)\psi]\end{align}
\begin{align}\Sigma_V&=(4\alpha^3\beta+2\alpha^2\beta^2-\alpha\beta)V-4\alpha^3\beta-2\alpha^2\beta^2+\alpha\beta
\notag
\\&r[(16\alpha^4+16\alpha^3\beta+4\alpha^2\beta^2+4\alpha^2\beta-16\alpha^2+2\alpha\beta)\psi\notag \\&
+(16\alpha^4+16\alpha^3\beta+4\alpha^2\beta^2-6\alpha^2
\beta+\alpha\beta^2-16\alpha^2+10\alpha\beta-\beta^2)f] \notag
\\& r^2[(-4\alpha^3-2\alpha^2\beta-4\alpha\beta-2\beta^2+4\alpha)\psi^2+(-8\alpha^4-24\alpha^3\beta-18\alpha^2\beta^2
\notag \\&-4\alpha\beta^3-11\alpha^2\beta-
4\alpha\beta^2+8\alpha^2+15\alpha\beta-2\beta^2)f\psi\notag
\\&+((8\alpha^4+8\alpha^3\beta+2\alpha^2\beta^2-8\alpha^2+\alpha\beta)p_r-\alpha\beta\rho
)V\notag
\\&+(-8\alpha^4-16\alpha^3\beta-18\alpha^2\beta^2-6\alpha\beta^3+8\alpha^2+17\alpha\beta-6\beta^2)f^2]\end{align}
\begin{align}\Sigma_f&=\psi[(8\alpha^4+4\alpha^3\beta-2\alpha^2)V+24\alpha^4+12\alpha^3\beta-14\alpha^2]\notag\\&
+f[(8\alpha^4-2\alpha^2\beta^2-2\alpha^2+
\alpha\beta)V+24\alpha^4-6\alpha^2\beta^2-14\alpha^2+7\alpha\beta]\notag\\&+r[f^2(-16\alpha^4-32\alpha^3\beta-20\alpha^2\beta^2-4\alpha\beta^3
-12\alpha^3+8\alpha^2\beta-\alpha\beta^2\notag\\&+28\alpha^2+8\alpha\beta+\beta^2
)+\psi
f(-16\alpha^4-24\alpha^3\beta-8\alpha^2\beta^2+4V\alpha\beta-2V\beta^2\notag\\&+4\alpha^3-2\alpha^2\beta-2\alpha\beta^2+40\alpha^2+8\alpha\beta)
+(16\alpha^4+8\alpha^3\beta-8\alpha^2)Vp_r\notag
\\&+4V\alpha\beta\psi^2+(8\alpha^3+12\alpha^2+4\alpha)\psi^2]+r^2[f^3(32\alpha^4+16\alpha^3\beta-8\alpha^2\beta^2\notag\\&
-4\alpha\beta^3-14\alpha^2-29\alpha\beta
+6\beta^2)+f^2\psi(32\alpha^4+24\alpha^3\beta-4\alpha^2\beta^2\notag\\&-4\alpha\beta^3-2V\alpha\beta-5V\beta^2-26\alpha^3-9\alpha^2\beta
-\alpha\beta^2-20\alpha^2-41\alpha\beta+2\beta^2)\notag\\&f\psi^2(-2V\alpha\beta-4V\beta^2-18\alpha^3-18\alpha^2\beta-6\alpha\beta^2-14\alpha^2
-12\alpha\beta+2\beta^2\notag\\&-4\alpha)+fV\rho(-2\alpha^2+\alpha\beta)+fVp_r(-8\alpha^4+2\alpha^2\beta^2+10\alpha^2-\alpha\beta)]\notag\\&\end{align}
\begin{align}\Sigma_\psi&=4\alpha^2\beta(1-V)+r[(8\alpha^2\beta^2+20\alpha^3-26\alpha^2\beta-9\alpha\beta)
\psi+\psi
V(12\alpha^3+2\alpha^2\beta\notag\\&-2\alpha\beta^2+\alpha\beta)+2\psi
V^3\alpha\beta^2
+f\psi(24\alpha^3\beta-4\alpha^2\beta^2+20\alpha^3-54\alpha^2\beta+2\alpha\beta^2)\notag\\&
+fV\psi(12\alpha^3-2\alpha^2\beta+2\alpha\beta^2)]+r^2[(16\alpha^3-12\alpha^2\beta)Vp_r+(-24\alpha^4-8\alpha^3\beta\notag\\&+14\alpha^2\beta^2
-2\alpha\beta^3+24\alpha^3+4\alpha^2\beta
+26\alpha\beta^2+2\beta^3)f^2
+fV\psi(8\alpha^2\beta+4\alpha\beta^2)\notag\\&+(-24\alpha^4+20\alpha^3\beta+10\alpha^2\beta^2+4\alpha\beta^3+48\alpha^3
+46\alpha^2\beta+17
\alpha\beta^2+32\alpha^2\notag\\&-6\alpha\beta+\beta^2)\psi f+
(8\alpha^2\beta+4\alpha\beta^2+2\beta^2)\psi^2V+4\alpha^2\beta\rho
V+(-8\alpha^3\beta-4\alpha^2\beta^2\notag\\&+24\alpha^3+28\alpha^2\beta+10\alpha\beta^2+24\alpha^2+2\alpha\beta)\psi^2]+
r^3[(4\alpha^2\beta+2\alpha\beta^2)V^2\psi
p_r\notag\\&+(12\alpha^3+6\alpha^2\beta+8\alpha^2-\alpha\beta)V\psi
p_r+(4\alpha^3+18\alpha^2\beta-2\alpha\beta^2)Vfp_r\notag\\&+(-2V\alpha\beta-16\alpha^3-16\alpha^2\beta-4\alpha\beta^2+4\alpha^2+4\alpha\beta
+2\beta^2-4\alpha
)\psi^3\notag\\&+(-36\alpha^4-30\alpha^3\beta-6\alpha^2\beta^2-28\alpha^3-37\alpha^2\beta-5\alpha\beta^2+4\beta^3-8\alpha^2\notag\\&
-25\alpha\beta+2\beta^2)f\psi^2+
(-4\alpha^2\beta-10\alpha\beta^2-4\beta^3-\beta^2)Vf\psi^2\notag\\&+(-36\alpha^4-30\alpha^3\beta+12\alpha^2\beta^2+6\alpha\beta^3+24\alpha^3
-76\alpha^2\beta
-58\alpha\beta^2\notag\\&+4\beta^3-16\alpha^2-17\alpha\beta+6\beta^2
)f^2\psi+(-4\alpha^2\beta-6\alpha\beta^2-6\beta^3)f^2\psi
V\notag\\&+(-4\alpha^3-2\alpha^2\beta+\alpha\beta)\rho\psi
V+(36\alpha^3-54\alpha^2\beta-46\alpha\beta^2+12\beta^3)f^3\notag\\&+(-4\alpha^3-2\alpha^2\beta+2\alpha\beta^2)\rho
Vf]\end{align} and
\begin{align}\Sigma_{p_t}&=(16\alpha^5+16\alpha^4\beta+4\alpha^3\beta^2-4\alpha^3-2\alpha^2\beta)V\psi\notag\\&
+(48\alpha^5-12\alpha^3\beta^2-28\alpha^3+10\alpha^2\beta
)\psi\notag\\&+(16\alpha^5+20\alpha^4\beta+6\alpha^3\beta^2-4\alpha^3-3\alpha^2\beta)Vf\notag\\&+
(48\alpha^5-36\alpha^4\beta-30\alpha^3\beta^2-28\alpha^3+27\alpha^2\beta)f\notag\\&+
r[(64\alpha^5+64\alpha^4\beta+16\alpha^3\beta^2-24\alpha^4-14\alpha^3\beta+3\alpha^2\beta^2\notag\\&-40\alpha^3+50\alpha^2\beta-15\alpha\beta^2
)f^2+(8\alpha^2\beta-6\alpha\beta^2)V\psi
f\notag\\&+((8\alpha^2\beta-6\alpha\beta^2)V+10\alpha^2\beta-10\alpha\beta^2+160\alpha^4\beta+60\alpha^3\beta^2\notag\\&+4\alpha^2\beta^3
-4\alpha^2\beta^2+112\alpha^5+8\alpha^4-64\alpha^3)\psi
f\notag\\&+(8\alpha^2\beta-2\alpha\beta^2)V\psi^2+(48\alpha^5+80\alpha^4\beta+28\alpha^3\beta^2+16\alpha^4\notag\\&+12\alpha^3\beta
+2\alpha^2\beta^2-24\alpha^3-16\alpha^2\beta+8\alpha^2-4\alpha\beta
)\psi^2\notag\\&+(32\alpha^5+8\alpha^4\beta-4\alpha^3\beta^2-16\alpha^3+4\alpha^2\beta)Vp_r]
\notag\\&+r^2[(-4\alpha^3-2\alpha^2\beta)\psi
V\rho+(-4\alpha^3-3\alpha^2\beta)fV\rho+(4\alpha^2\beta-\alpha\beta^2)\psi
V^2p_r\notag\\&+(24\alpha^5+24\alpha^4\beta+6\alpha^3\beta^2+8\alpha^4+2\alpha^3\beta-\alpha^2\beta^2-12\alpha^3)\psi
Vp_r\notag\\&+(32\alpha^5+48\alpha^4\beta+12\alpha^3\beta^2-2\alpha^2\beta^3-28\alpha^3+3\alpha^2\beta)fVp_r\notag\\&
+(4\alpha\beta^2-2\alpha\beta+\beta^2)\psi^3V+(-4\alpha^2\beta+2\beta^3)f\psi^2
V\notag\\&+(-4\alpha^2\beta-6\alpha\beta^2+3\beta^3)f^2\psi
V\notag\\&+(-32\alpha^5-32\alpha^4\beta-8\alpha^3\beta^2-8\alpha^4+2\alpha^2\beta^2+12\alpha^3\notag\\&
-8\alpha^2\beta-3\alpha\beta^2+12\alpha^2+2\alpha\beta
)\psi^3\notag\\&+(-112\alpha^5-160\alpha^4\beta-60\alpha^3\beta^2-4\alpha^2\beta^3-64\alpha^4\notag\\&
-60\alpha^3\beta-6\alpha^2\beta^2+4\alpha\beta^3+84\alpha^3
+18\alpha^2\beta-6\alpha\beta^2+20\alpha^2
)f\psi^2\notag\\&+(-88\alpha^5-240\alpha^4\beta-166\alpha^3\beta^2-26\alpha^2\beta^3+4\alpha\beta^4-52\alpha^4\notag\\&
-65\alpha^3\beta-12\alpha^2\beta^2+3\alpha\beta^3
+112\alpha^3+75\alpha^2\beta-12\alpha\beta^2
)f^2\psi\notag\\&+(-8\alpha^5-104\alpha^4\beta-122\alpha^3\beta^2-24\alpha^2\beta^3\notag\\&+6\alpha\beta^4+44\alpha^3+51\alpha^2\beta-18\alpha\beta^2
)f^3].\end{align}

\begin{figure}
\centering\subfigure[{}]{\label{1011}
\includegraphics[width=0.5\textwidth]{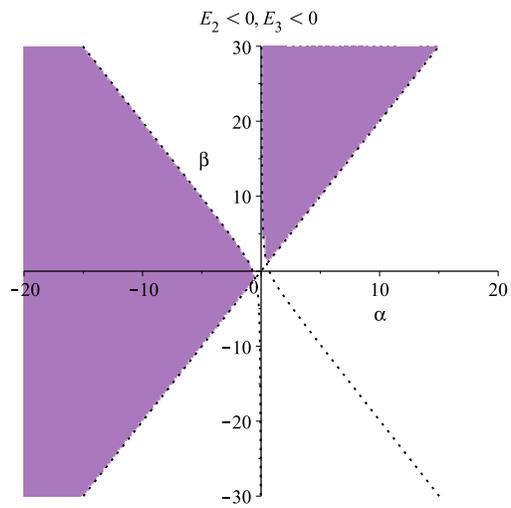}}
\hspace{2mm}\subfigure[{}]{\label{23411}
\includegraphics[width=0.5\textwidth]{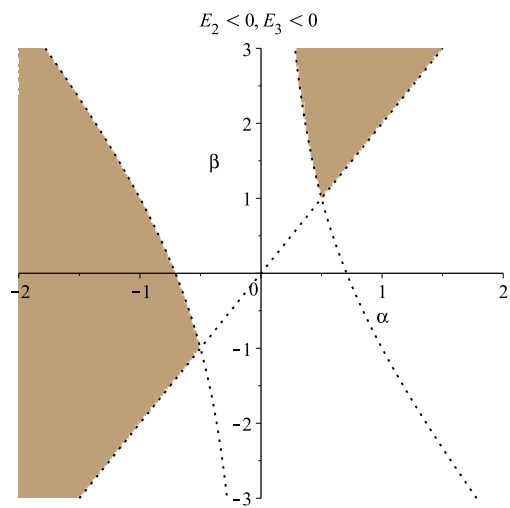}}
\hspace{2mm}\caption{\footnotesize{permissible values for $\alpha$
and $\beta$ parameters with stable state of the solutions
(negative eigenvalues $E_{2,3}<0$)}}
\end{figure}
\begin{figure}
\centering\subfigure[{}]{\label{1011}
\includegraphics[width=0.4\textwidth]{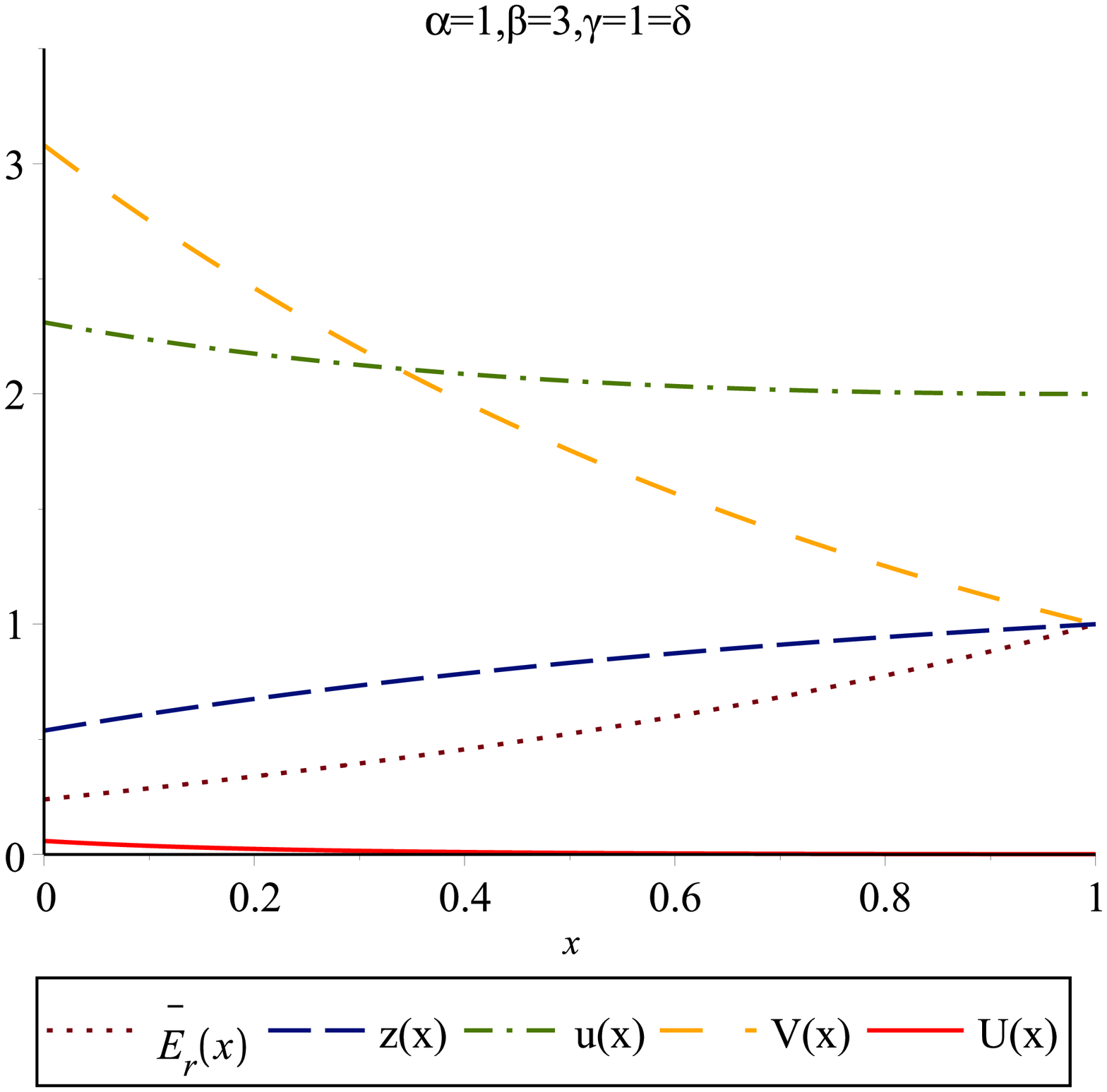}}
\hspace{2mm}\subfigure[{}]{\label{eigen}
\includegraphics[width=0.4\textwidth]{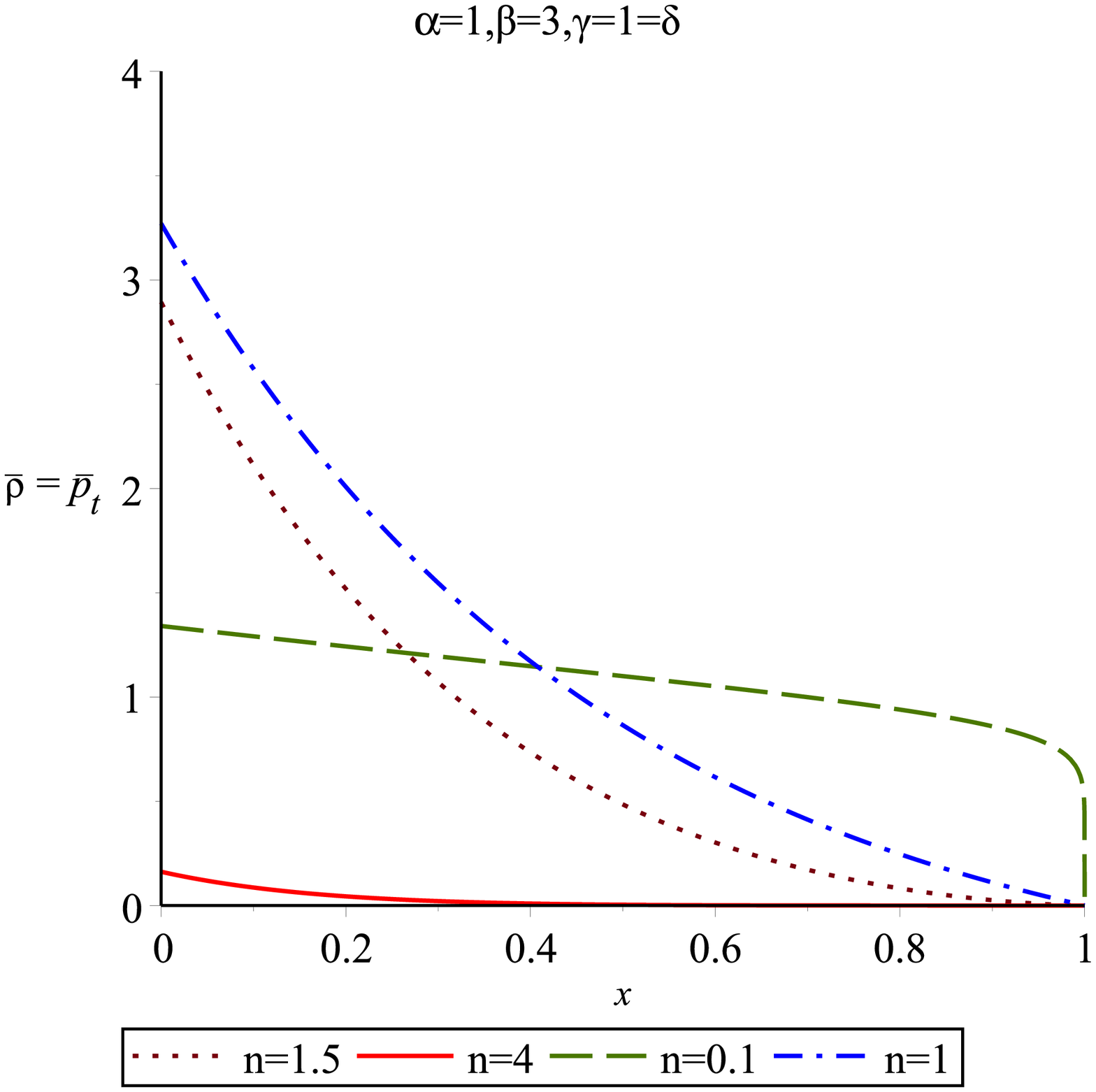}}
\hspace{2mm}\subfigure[{}]{\label{z1}
\includegraphics[width=0.4\textwidth]{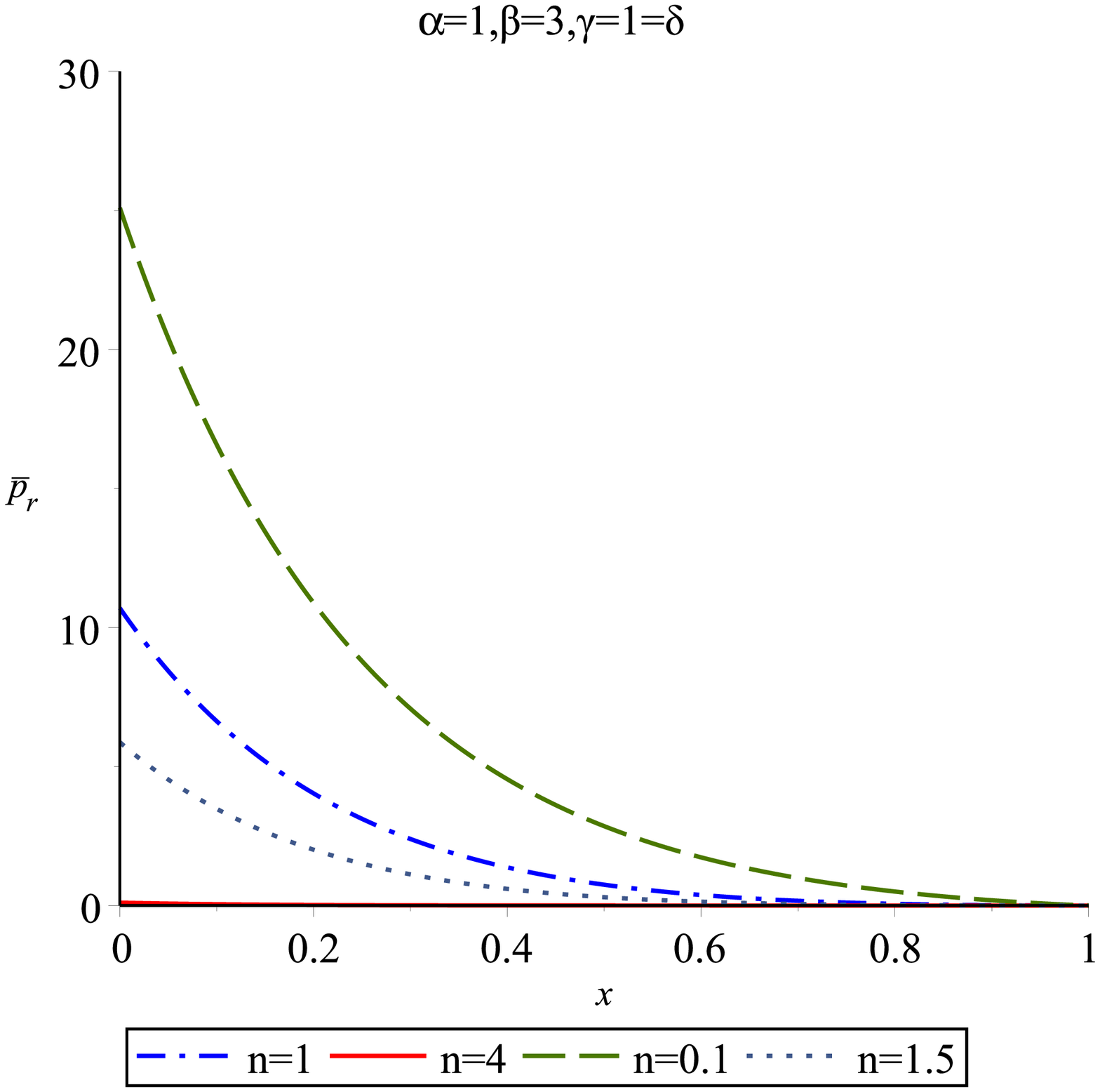}}
\hspace{2mm}\subfigure[{}]{\label{b1}
\includegraphics[width=0.4\textwidth]{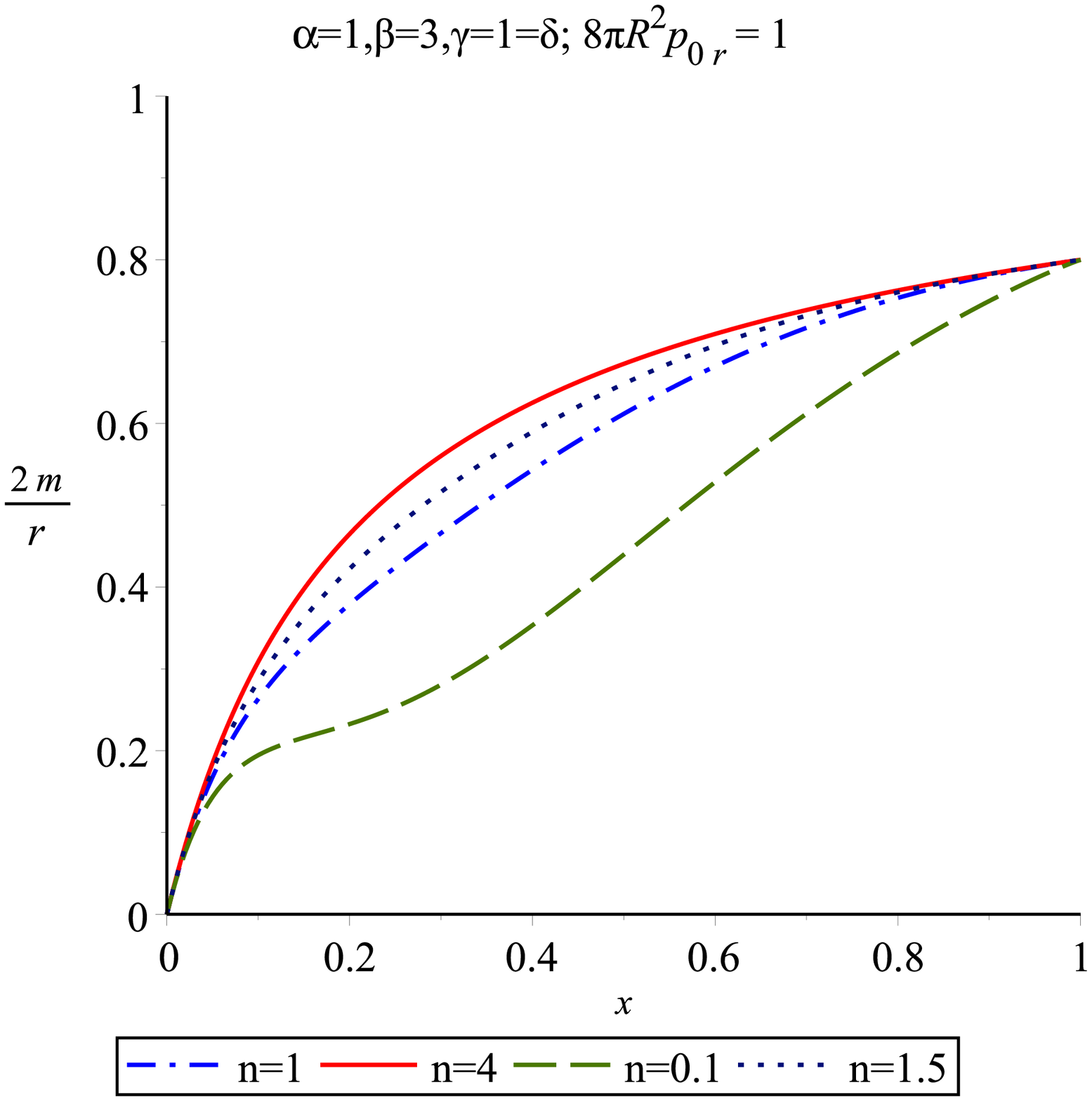}}
\hspace{2mm}\caption{\footnotesize{Diagrams of the fields for
sample permissible values of the parameters
$\alpha=1,\beta=3,\delta=\gamma=1$ }}
\end{figure}
\end{document}